\documentstyle[sprocl]{article}
\input{epsf}
\def\be{\begin{equation}}
\def\ee{\end{equation}}
\def\bea{\begin{eqnarray}}
\def\eea{\end{eqnarray}}

\newcommand{\NI}{\noindent}

\input{epsf}

\begin{document}

\title{MATCHING WEAK COUPLING AND QUASICLASSICAL EXPANSIONS FOR DUAL
QES PROBLEMS}

\author{MICHAEL KAVIC}

\address{Department of Physics,University of Minnesota Tate Laboratory
of Physics, \newline 116 Church Street S.E. Minneapolis, MN 55455, USA}

\author{}

\address{}

\maketitle{}
\abstracts{Certain quasi-exactly solvable systems
exhibit an energy reflection property that relates the energy levels of
a potential or of a pair of potentials. We investigate two sister 
potentials and show the existence of this energy reflection
relationship between the two potentials. We establish a relationship 
between the lowest energy edge in the first potential using the weak coupling expansion 
and the highest energy 
level in the sister potential using a WKB
approximation carried out to higher order. }

\section{Introduction}
The discovery of quasi-exactly solvable (QES) spectral problems [1-4] 
provided new opportunities for exploring such interesting issues as the convergence of perturbative series at high
orders or the relationship between weak coupling and quasiclassical (WKB) expansions.[5] Of special importance for the 
second issue is the energy reflection (ER) symmetry of certain QES potentials [5,6]. It was noted that in certain potentials the QES part of the spectrum is 
symmetric under $E \leftrightarrow -E$. The transformation rule of the corresponding wave functions is also known.

In this work we will limit ourselves to a particular example of the so-called
dual potentials [6]. One-dimensional QES problems are based on sl(2).
This is represented as follows,
\begin{equation}
 H \Rightarrow \sum_{a,b} C_{ab}T^{a}T^{b}+\sum_{a} C_{a}T^{a},
\end{equation}
\NI where $C$'s are constants and $T$'s are sl(2) generators,
\begin{eqnarray}
T^{+}=+2j\xi -\xi^{2}\frac{d}{d\xi},
			\nonumber\\
T^{0}=-j +\xi \frac{d}{d\xi},\quad\,\,\,\,
			\\
T^{-}=+\frac{d}{d\xi}.\quad\quad\quad\,\,\, 
			\nonumber
\end{eqnarray}
\NI The generators act on polynomials of $\xi$ of degree $2j$ provided  $j$ is a semi-integer
number. The dual potentials [6] are such that the two sister potential are related by the
mapping,       
\begin{equation}
E \rightarrow -E.
\end{equation}
\NI Quasi-exact solutions in quantum-mechanics problems depend on the cohomology parameter
$(j)$. This parameter determines how much of system's
spectral levels are included in the algebraic sector, or, in other words,
how large the $2j \times 2j$ matrix will be [7]. 

An interesting limit may be taken,
$ j \rightarrow \infty $. If  $V(x)$ is a potential well, 
it may become deeper as $j$ is increased. If $ j \rightarrow \infty $ the lowest 
energy level can be calculated perturbatively since the anharmonicity becomes small. 
This is done by performing an expansion
about the minimum point, approximating the potential at the minimum as a 
slightly perturbed harmonic oscillator.

This expansion leads to an expression for the ground state energy. In the dual potential 
the energy levels are symmetric with respect to zero. So, the energy 
reflection relationship can lead to an expression for the highest QES energy level
of its sister. This expression may be checked by more 
traditional means. In particular, a WKB approximation can be applied for highly excited states.
Each term is calculated using higher order
WKB approximations. The WKB expansion can be confronted with the weak coupling 
expansion for the ground state of the sister potential. They should match. In Ref [5]
a self dual potential was studied in this context. Here I analyze a pair of dual potentials
with the aim of establishing the matching. 

\section{Lowest Energy Calculation}
Let us investigate a particular example.
\begin{equation}
	V_1(x) = \frac{1}{2}\sin^{2} x + \left(2j+\frac{1}{2}\right)\cos x, 
\end{equation}
\begin{equation}
	V_2(x) = \frac{1}{2} \sinh^{2} x - \left( 2j+\frac{1}{2} \right) \cosh x.      
 \end{equation}
\begin{figure}[h]
\begin{center}
\epsffile{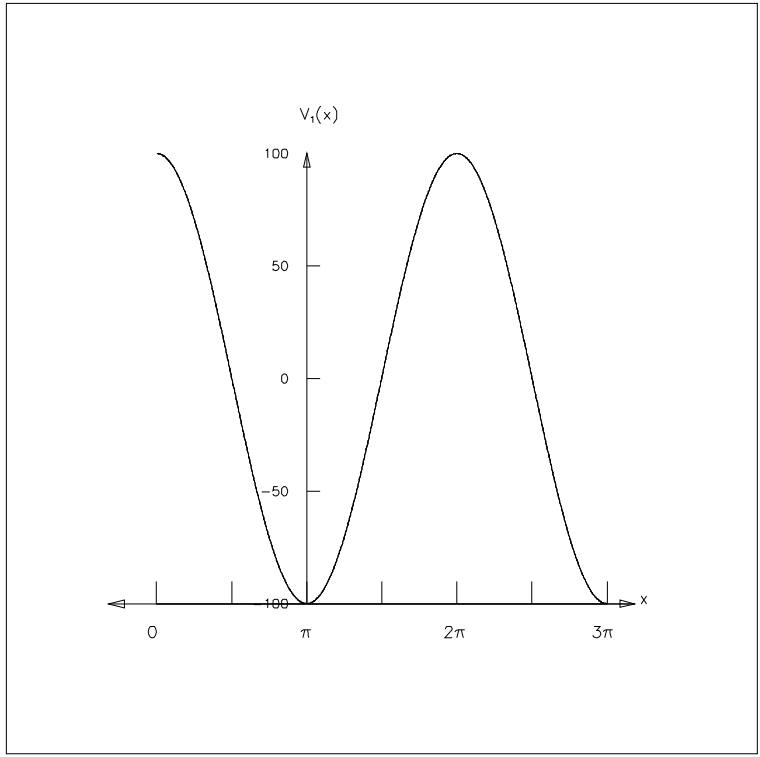}
\caption{$V_1(x),\quad  \kappa=101$}
\label{fig:potential_1}
\end{center}
\end{figure}
\begin{figure}
\begin{center}
\epsffile{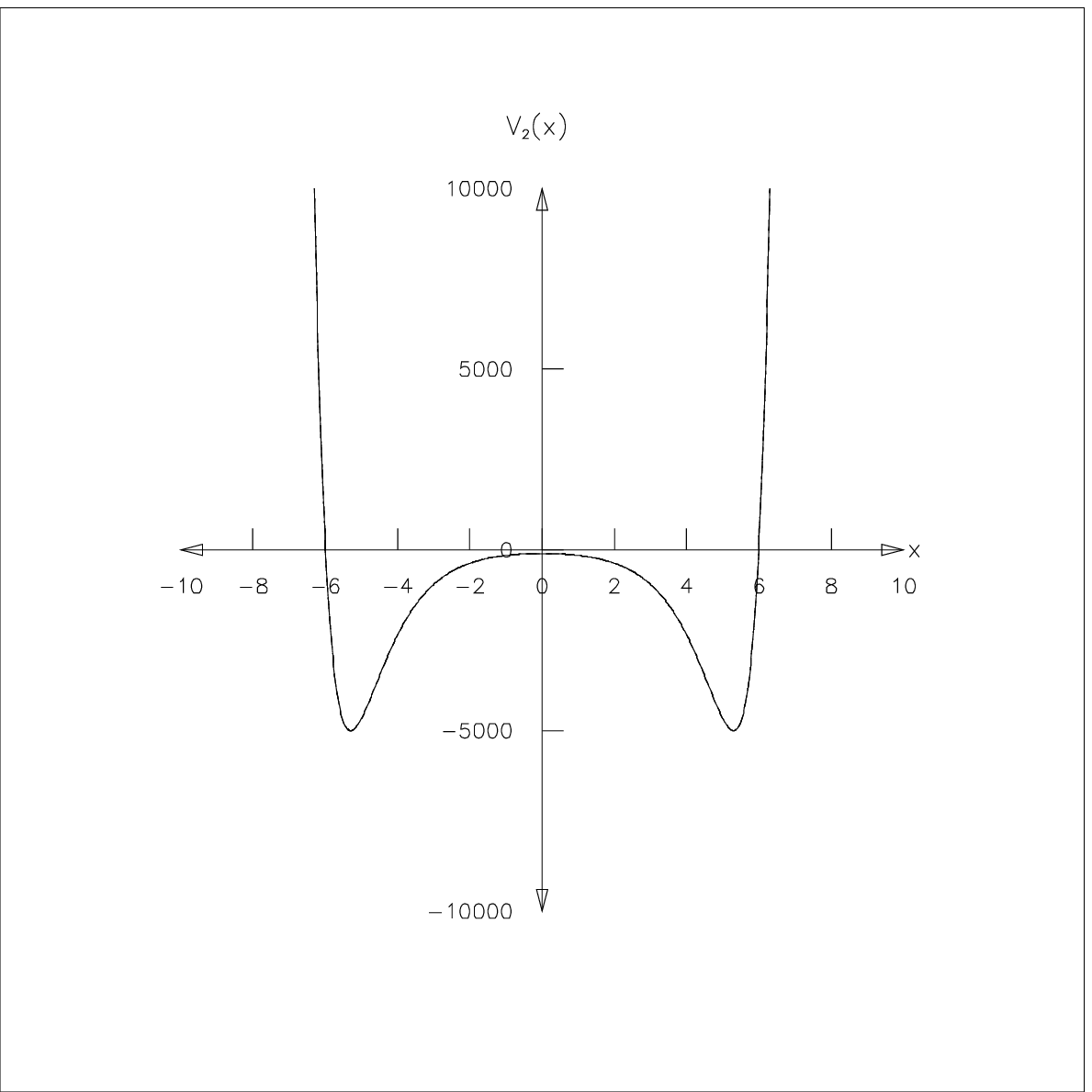}
\caption{$V_2(x),\quad  \kappa=101$}
\label{fig:potential_2}
\end{center}
\end{figure}

\NI The potentials are plotted in Figures 1 and 2. We start from the weak coupling expansion for
$V_{1}(x)$.  The first step is to calculate an
expression for $E_{0}$. Because the potential is periodic the expression
corresponds to the lower edge of the lowest energy band.

Let us parenthetically note that at large $j$ the width of the band can be readily evaluated
quasiclassically by calculating the tunneling amplitude from $x=\pi$ to 
$x=3\pi$,
\begin{equation}
S_{\rm tunnel}=8\sqrt{2j+\frac{3}{2}}, 
\end{equation}
\NI and the band width 
\begin{equation}
\delta {E} \propto
e^{-S_{\rm tunnel}}.
\end{equation}
We neglect the effects that arise due to tunneling because they
affect only terms exponentially small in $j$ while we address here
the power expansion in $\frac{1}{\sqrt{j}}$.

As previously suggested, we consider the limit $j \gg 1$. The eigenvalue 
problem takes the form
\begin{equation}
  \left[-\frac{1}{2}
  \frac{d^{2}}{dx^{2}}+
 [\frac{1}{2}\sin^{2} x + 
 \gamma\cos x]\right]\psi(x) = E\,\psi(x),
  \end{equation}
\NI where,
\begin{equation}
  \gamma=2j+\frac{1}{2}, \quad \textrm{($j$ is semi-interger)}.
\end{equation} 
\NI Taking the limit $\gamma \rightarrow \infty$ causes the potential well to become
deeper. The minimum can be calculated to lie at, 
\begin{equation}
 x_{\textrm{\scriptsize{min}}}=\pi.         
\end{equation}
 \NI Expanding about $ \pi $  and using the following notation,
\begin{equation}
\kappa=(\gamma+1)
\end{equation}
\NI gives the following expression for $V_1(x)$:
\begin{equation}
V_1(x)=-(\kappa-1)+\frac{1}{2}\sqrt{\kappa}(x-\pi)^{2}-	
	\frac{1}{24} ( \kappa+3) (x-\pi)^{4}+\frac{1}{720}(\kappa+15)(x-\pi)^{6}+\cdots,
\end{equation}
where the dots denote higher order terms.

The terms of 
the $\frac{1}{\sqrt{\kappa}}$ expansion can be determined by treating the
 potential at the minimum as a slightly perturbed harmonic oscillator. 
The leading term is the classical term,
\begin{equation}
E_{0}=- (\kappa-1).
\end{equation}
\NI The linear in $(x-\pi)$ term in $V_{1}$ vanishes as do all the odd-power terms. 
The next term in $E_{0}$ comes from the zero-point oscillation energy for the 
a harmonic oscillator,
\begin{equation}
\delta E_{0}=\frac{\omega}{2} = \frac{1}{2}\sqrt{\kappa}.
\end{equation}
\NI Next we consider the anharmonic terms. The first order perturbation in $(x-\pi)^{4}$
produces the correction $\propto\kappa^{0}$,
\begin{equation}
\Delta_{1} E_{0}=-\frac{1}{32}\,. 
\end{equation}
\NI The following term is given by adding the second order perturbation of the quartic term,
\begin{equation}
\Delta_{2} E_{0}=-\frac{7}{64}\frac{1}{\sqrt{\kappa}}, 
\end{equation}
with the first order perturbation of the sextic term\,,
\begin{equation}
\widetilde{\Delta_{2}E_{0}}=\frac{1}{384}\frac{1}{\sqrt{\kappa}}\,.
\end{equation}
\NI Taken together these terms give the following $E_{0}$ expansion: 
\begin{equation}
E_{0}=-\kappa+\frac{1}{2}\sqrt{\kappa}+\frac{31}{32}
-\frac{41}{384}\frac{1}{\sqrt{\kappa}}\cdots\,.
\end{equation}
\section{WKB Calculation}
If the energy levels of the sister potentials are symmetric with respect to zero, then
multiplying by $(-1)$ the lowest energy expansion should provide an expression for the highest energy level in the algebraic sector
of the sister potential. So, the energy reflection property takes the form
\begin{equation}
-E_{0}=E_{*}\,,
\end{equation}
\NI where $E_{*}$ is the highest energy level in the sister
potential which is linked to the original potential by mapping $V(x)\rightarrow -V(xi)$.
Taking this as the highest even QES energy level in the potential depicted
in Fig.2, we determine the number of this (P-even) state in terms of $\gamma$,
\begin{equation}
n=2(\gamma-\frac{1}{2}).
\end{equation}
\NI Thus, from our calculation of the lowest energy in $V_{1}$ we have 
determined an expression
for the highest energy in the sister system $V_{2}$. We now
use WKB approximations to confirm this expression. Our second potential is 
\begin{equation}
 V_{2}(x)=\frac{1}{2} \sinh^{2} x - 
\gamma \cosh x. 
\end{equation}
\NI
We scale the energy 
in the Schr{\"{o}}dinger equation such that $\alpha$ is the scaled energy,
$E\equiv\gamma\alpha$.
We use the convenient change of variable $x=2t$. Thus, 
\begin{equation}
[E-V_{2}(t)]=(\gamma\alpha-\sinh^{2} t)\,2\cosh^{2}t.
\end{equation}
\NI  
After scaling the independent variables 
we obtain the Schr{\"{o}}dinger equation,
\begin{equation}
 \epsilon\,\psi^{''}(t)=[(1-\frac{1}{\gamma}\sinh^{2} t)\,2\cosh^{2}t-
(1-\alpha)]\,\psi (t),
\end{equation}
\NI where the small parameter is defined as $\epsilon=\frac{1}{2 \sqrt{2\gamma}} $. Adopting the convention $\beta=(1-\alpha)$, the scaled energy expansion becomes
\begin{equation}
\beta=\sum_{n=0}^{\infty}a_{n} \epsilon^{n}\,.
\end{equation} 
\NI We will now use WKB approximation carried out to higher order to determine the coefficients of the energy expansion ($a_{n}$). The WKB quantization condition [8,9] to $\epsilon^{5}$ is
\begin{equation}
2(\gamma-\frac{1}{2})\pi=\frac{1}{2 \epsilon}\oint dt \sqrt{Q}-\frac{\epsilon}{96}\oint dt 
\frac{Q^{''}}{Q^{3/2}}-\frac{\epsilon^{3}}{3072}\oint dt \frac{2Q^{''''}Q-7(Q^{''})^{2}}{Q^{7/2}}+\cdots,
\end{equation}
\NI in which $Q(t)=2[(\gamma-\sinh^{2} t)\,2\cosh^{2} t]$ and the contours encircle 
the two turning points with the branch 
cut joining them. Once again the dots refer to higher order
terms.  

Next we expand Eq. (25) in powers of $\epsilon$ and evaluate the 
contour integrals. In the 
calculation of each integral we use a second change of variables $y=\sinh t $. 
The first integral can be done along the real axis. The following integrals
need to be evaluated along a contour but can be done in closed form. 
Each of the integral can be evaluated as a multiple of $\pi$. The factor 
of $\pi$ is then canceled with the factor of $\pi$ in the 
quantization condition, leaving a purely algebraic 
series. 

Using the results from Eq. (25) we can calculate the scaled energy 
coefficients ($a_{n}$) in Eq. (24). This results in the 
following energy expansion
\begin{equation}
E_{*}=\gamma-\frac{1}{2}\sqrt{\gamma+1}+\frac{1}{32}
+\frac{41}{384}\frac{1}{\sqrt{\gamma+1}}\cdots\,.
\end{equation}
\NI Using again the notation $\kappa=\gamma+1$ yields
\begin{equation}
E_{*}=\kappa-\frac{1}{2}\sqrt{\kappa}-\frac{31}{32}
+\frac{41}{384}\frac{1}{\sqrt{\kappa}}\cdots\,,
\end{equation}
in full agreement with the prediction given by the energy reflection relation.

\section{Conclusion}
The $\frac{1}{\sqrt{\kappa}}$ WKB expansion for $V_{2}$ is the same as the weak coupling
expansion for $V_{1}$, predicted by the energy reflection relation. Thus we have validated the
relation and established the relationship between the energy levels of the two dual 
potentials. We have shown not only that the relationship between the energy levels 
exist, but that it persits to higher
orders algebraically. This is remarkable because the pertubation calculation and the
 WKB approximation are independent
of one another. Thus, a deeper principle must underlie the symmetry
that exists between the energy levels of the two potentials.

\section{Acknowledgements}
M.K. thanks Professor Mikhail Shifman for his careful guidance
and generous support, and he thanks Xin Rui Hou, Justin Hietala, 
and Steven Giron for their shared knowledge and patient assistance.

\end{document}